\documentclass[aps,prb,twocolumn,superscriptaddress,longbibliography]{revtex4-2}
\usepackage[colorlinks=true,linkcolor=blue,citecolor=blue]{hyperref}
\usepackage{amsmath,amssymb}
\usepackage[utf8]{inputenc}
\usepackage[T1]{fontenc}
\usepackage{url}
\usepackage{adjustbox}
\usepackage{subcaption}
\usepackage{graphicx}
\usepackage{bm}
\usepackage{bbm}
\usepackage{xcolor}
\usepackage{color}
\usepackage{ulem}
\usepackage{subfloat}
\usepackage{ragged2e}
\usepackage{caption}
\usepackage{braket}
\DeclareCaptionFormat{myformat}{\justifying#1#2#3}
\newcommand{\bs}{\boldsymbol}
\graphicspath{{figures/}}

\begin{document}

\title{Transport properties in a two-dimensional Su-Schrieffer-Heeger model in Quantum Hall Regime}

\author{Aruna Gupta}
\email{arunagupta169@gmail.com}
\affiliation{Department of Physics, Birla Institute of Technology and Science, Pilani 333031, India} 

\author{Shaina Gandhi}
\email{p20200058@pilani.bits-pilani.ac.in}
\affiliation{Department of Physics, Birla Institute of Technology and Science, Pilani 333031, India} 

\author{Niladri Sarkar}
\email{niladri@pilani.bits-pilani.ac.in}
\affiliation{Department of Physics, Birla Institute of Technology and Science, Pilani 333031, India} 

\author{Jayendra N. Bandyopadhyay}
\email{jnbandyo@gmail.com}
\affiliation{Department of Physics, Birla Institute of Technology and Science, Pilani 333031, India}

\begin{abstract}
 
We investigate the transport properties of a two-dimensional Su-Schrieffer-Heeger (2D SSH) model in the quantum Hall regime using non-equilibrium Green's function formalism (NEGF). The device Hamiltonian, where the 2D SSH model serves as the channel, is constructed using a nearest-neighbor tight-binding model. The effect of an external perpendicular magnetic field is incorporated into the contacts via Peierls substitution. We observe a transition from a gapped phase to a flat band regime at zero energy by varying the magnetic field. This transition is characterized by the emergence of highly localized states in the bulk or edges, which we observe by calculating local density-of-states (LDOS). We analyze transport in the system along two directions ($x$ and $y$) via transmission measurements, indicating a magnetic field-induced transition from insulating to metallic phase. The study of the energy spectrum of the system shows the formation of Landau levels. Moreover, the quantum number of the non-degenerate and degenerate Landau levels (transmission modes) can be any integer or only an odd integer, depending on diagonal, inter-cell, and intra-cell hopping strengths. From the analysis of the transport properties along $y$-direction, we find that edge modes play a crucial role in facilitating ballistic transport.
 
\end{abstract}

\maketitle 
\section{Introduction}

Topological insulators (TIs) are an exotic class of materials that are characterized by insulating bulk and conducting edges or surfaces. The edge states in TIs are remarkably robust, maintaining their conductivity even in the presence of disorder and perturbations. The discovery of the quantum Hall effect gives prominence to the TIs in condensed matter literature. It has shown that a two-dimensional (2D) metal in a strong magnetic field confines the motion of electrons in the bulk but forces them into delocalized at the edges (for 2D TIs) or at the surfaces (for 3D TIs). 

A paradigmatic example of the TIs is the Su-Schrieffer-Heeger (SSH) model \cite{PhysRevLett.42.1698}. This model is a one-dimensional (1D) model, represented by a chain, and it describes electrons hopping on the chain with intracell hopping strength $v$ and intercell hopping strength $w$. When $|v| < |w|$, the SSH model is topologically trivial, while when $|v| > |w|$, this model becomes topologically nontrivial and exhibits edge states in the system. Since this is a 1D model, the edge states reside near the endpoints. The SSH model's simplicity and versatility have led to its extensive exploration by experimental and theoretical means in various physical contexts like photonic, acoustic, and optical systems \cite{PhysRevApplied.20.034028, Khanikaev2013}, ultracold atoms \cite{Xie2019, PhysRevA.97.023618}, ferromagnetic systems \cite{PhysRevB.103.014407}, superconducting systems \cite{PhysRevB.107.235113, PhysRevB.106.054511}, electrical circuits \cite{PhysRevResearch.3.023056}, etc. Various modifications and extensions to the SSH model have also been widely studied, having long-range hopping terms in the SSH chain \cite{PhysRevB.99.035146}, two coupled SSH chains or ladder systems \cite{Zurita, PhysRevB.106.205111, Jangjan2020, Hetenyi_2018}, 2D extensions \cite{PhysRevB.108.104101}, and boundary induced effects \cite{PhysRevB.109.075106}. All these modifications impact the system's topology and edge states.

The discovery of the quantum Hall effect (QHE) in 1980 marked a significant milestone in condensed matter physics \cite{PhysRevLett.45.494}. Since its discovery, the QHE has been extensively studied in various systems, particularly in topological insulators \cite{KLITZING1983682}, where it has provided deep insights into the behavior of electrons in low-dimensional systems and the role of topology in quantum systems. Research has expanded beyond the conventional integer quantum Hall effect to include the fractional quantum Hall effect  \cite{rao1999fractionalquantumHalleffect}. More recently, unconventional quantum Hall effect has been observed in topological materials like graphene, Weyl semimetals, and other 2D materials \cite{PhysRevLett.96.086805, 1327329}. Furthermore, 2D materials like graphene have shown quantum transport phenomena, including the unconventional quantum Hall effect (QHE). In graphene, when subjected to an external magnetic field, the Hall resistance ($R_H$) forms plateaus at values such as \( h/(2e^2) \), \( h/(6e^2) \), \( h/(10e^2) \), and so forth, corresponding to two electronic valleys and two spin states \cite{PhysRevLett.95.146801}. This means that the Hall conductivity (\( \sigma_H \)) increases in steps related to odd integer multiples of \( e^2/h \) rather than the even integer steps seen in more conventional systems. Extensive theoretical work has been conducted to uncover the origin of this unusual sequence in the Hall plateaus observed in graphene. These studies have explained this phenomenon using a continuum model of edge states. They reveal that graphene's odd-integer QHE arises from a zeroth order Landau Level (LL) and two-fold valley-degenerate LLs. This unique Landau level structure is responsible for the unconventional quantum Hall effect seen in graphene \cite{PhysRevB.73.195411, PhysRevLett.96.176803}. This study observes a similar unconventional quantum Hall effect in the presence of the external magnetic field and specific hopping parameters in the 2D SSH model case. 

Building on these ideas, we investigate the transport properties of a 2D SSH model in the presence of a magnetic field. The model is constructed by stacking 1D SSH chains along the $y$-axis and introducing diagonal hopping between the two layers from A sublattice of one layer to B sublattice of another layer and vice versa) as shown in Fig. \ref{Schematic}. Our primary objective is to explore the formation of Landau levels in the quantum Hall regime and to understand how the magnetic field influences the conduction properties in the 2D SSH model. The zeroth LL generates an extra conducting edge mode by splitting in energy, moving up for electrons and down for holes near the edges when a hard wall confining potential is present \cite{article}. Our analysis examines the transport properties of the 2D SSH model in two distinct directions, which is analogous to the study of transport along the armchair and zigzag edges in graphene. We study transport properties by varying the diagonal hopping parameter \( t_d \) from $0$ to $1$, revealing intriguing results, particularly around \( t_d = 0.5 \) eV. We perform this study for the topologically trivial and the nontrivial cases, where each chain is in a topologically trivial or nontrivial phase depending on the values of $v$ and $w$, providing a comprehensive understanding of the system's behavior under different conditions. 

We organize this paper as follows. In Sec. \ref{Theory}, we describe a 2D SSH model and provide an analytical investigation of the \(E\)-\(k\) relation. This section also explains how the magnetic field is incorporated into the channels and the contacts using the Peierls substitution. Next, in Subsection \ref{Tp}, we describe the calculation of transport properties using the NEGF formalism, along with the corresponding band structures and local density of states to analytically verify transmission (conductance). In Sec. \ref{results}, we present an in-depth discussion of the results, highlighting the interplay between the diagonal hopping parameter and charge conductance. Finally, in Sec. \ref{summary}, we summarise our findings briefly.

\section{Model}
\label{Theory}

\begin{figure}	
    \includegraphics[width=9cm,height=6cm]{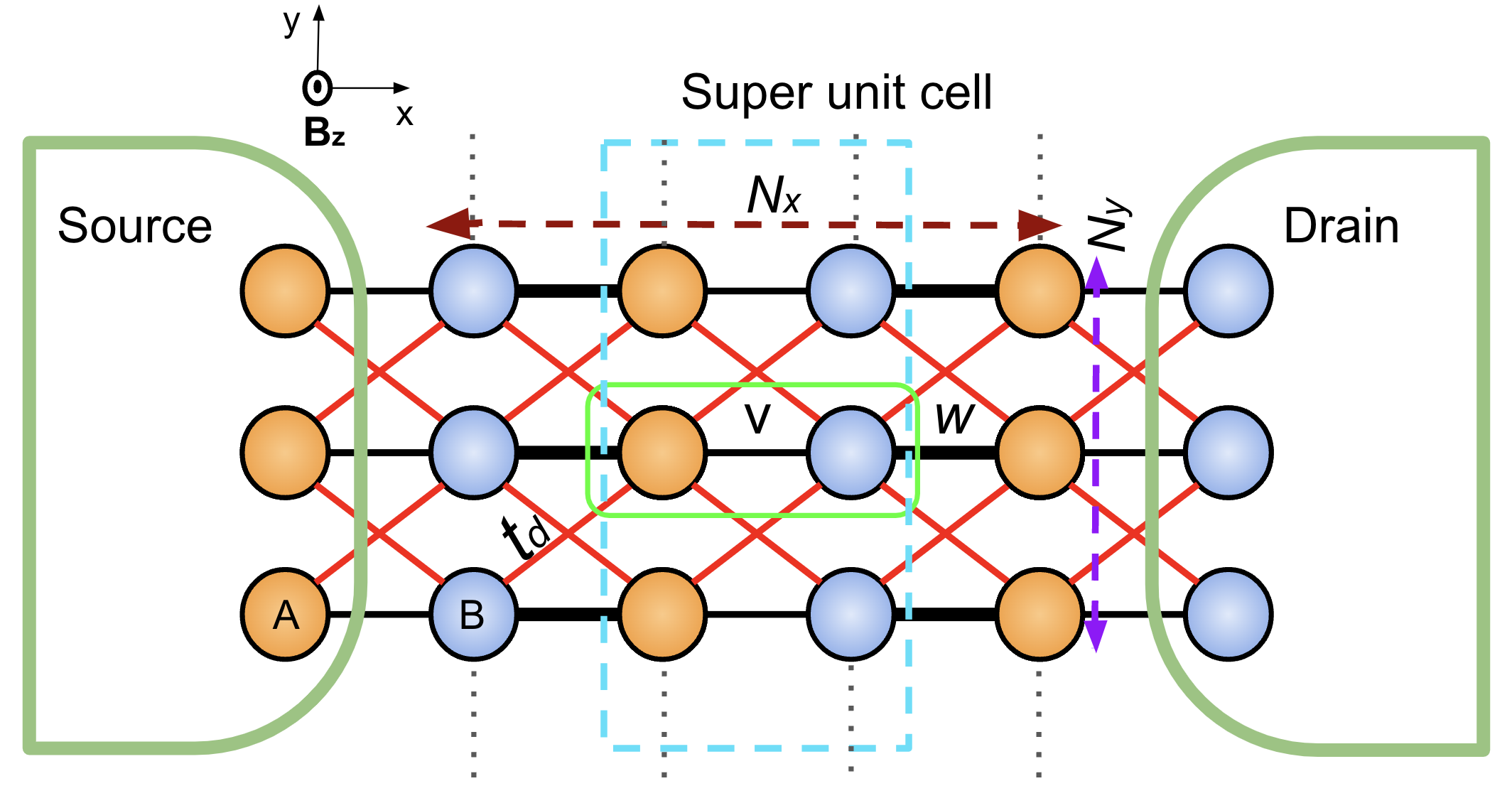}
 \caption{A schematic diagram of the tripartite system is presented. The red-colored circle represents sublattice  $A$, and the blue-colored circle represents sublattice $B$ of a unit cell. A green-colored rectangular box shows a unit cell. The black thin line corresponds to intercell hopping $v$ while the black thick line corresponds to intracell hopping $w$. The red-solid lines represent the diagonal hopping of strength $t_d$ between the consecutive layers from sublattice $A$ to sublattice $B$. The ends of the 2D SSH channel are connected to the source and drain leads.}
\label{Schematic}
\end{figure}

In this tripartite system, as illustrated in Fig. \ref{Schematic}, the setup includes a large system described by a 2D SSH model as the central channel placed between a drain and a source along $x$-direction. A detailed description of this setup is discussed in Ref. \cite{PhysRevA.95.062114, article1}. This channel is visualized as an array of supercells comprising multiple transversely arranged unit cells. The width of the channel $N_y$ is defined by the number of dimers in the transverse direction, while the length $N_x$ is determined by the number of dimers along the longitudinal direction. The ends of the 2D SSH channel are connected to the source and drain leads, with the channel described by the Hamiltonian \(H_{\rm SSH}\). The coupling to the leads is represented by the broadening term \(\Gamma\), connecting to the left lead (source) and right lead (drain), with their respective self-energy Hamiltonians \(\Sigma_1\) and \(\Sigma_2\). The total Hamiltonian of the system is given by:
\begin{equation}
 H = H_{\rm SSH} + \Sigma_1 + \Sigma_2 
 \label{tot_Ham}
\end{equation}
The real-space Hamiltonian for the channel is:
\begin{equation}
\begin{aligned}
H_{SSH} = & -\sum_{n_{x},n_{y}} \left( v \, c^{A^{\dagger}}_{n_{x}n_{y}} c^{B}_{n_{x}n_{y}} + w \, c^{B^{\dagger}}_{n_{x}n_{y}} c^{A}_{n_{x}n_{y}} + \text{H.C.} \right) \\
& + \sum_{n_{x},n_{y}} \left( t_d \, c^{A^{\dagger}}_{n_{x}n_{y}} c^{B}_{n_{x},n_{y}\pm 1} + t_d \, c^{B^{\dagger}}_{n_{x}n_{y}} c^{A}_{n_{x},n_{y}\pm 1} \right)
\end{aligned}
\label{SSH_Ham}
\end{equation}
The Hamiltonians for the left and right leads are, respectively:
\begin{equation}
\Sigma_1 = \sum_{n_{x},n_{y}} \left( l_{n_{x},n_{y}}^\dagger l_{n_{x}+1,n_{y}+1} + l_{n_{x}+1,n_{y}+1}^\dagger l_{n_{x},n_{y}} \right)
\end{equation}
\begin{equation}
\Sigma_2 = \sum_{n_{x},n_{y}} \left( r_{n_{x},n_{y}}^\dagger r_{n_{x}+1,n_{y}+1} + r_{n_{x}+1,n_{y}+1}^\dagger r_{n_{x},n_{y}} \right)
\end{equation}
Here, \({n_{x}, n_{y}}\) are site indices, and \(\{c_{n_{x}, n_{y}}^\dagger, c_{n_{x}, n_{y}}\}\), \(\{l_{n_{x}, n_{y}}^\dagger, l_{n_{x}, n_{y}}\}\), \(\{r_{n_{x}, n_{y}}^\dagger, r_{n_{x}, n_{y}}\}\) represent the creation and destruction operators for the sites of the SSH chain, left lead, and right lead, respectively. The self-energy matrix \(\Sigma\) modifies the Hamiltonian \([H]\) to incorporate the boundary conditions.

This study considers a unit cell containing two lattice sites labeled $A$ and $B$. The 2D SSH model is constructed with $N_y = 60$ unit cells in the transverse or $y$-direction and four supercells in the longitudinal or $x$-direction. Each supercell comprises $120$ lattice sites, leading to $480$ within the system. The intracell hopping parameter within each unit cell is denoted by \(v\), while hopping between adjacent unit cells, or intercell hopping, is denoted by \(w\). Additionally, we introduce an interlayer diagonal hopping parameter, \(t_d\), which represents hopping from a site $A$ in one layer to a site $B$ in the adjacent layer.

\subsection{$\mathbf{E-k}$ relation}
\label{EK}

The Hamiltonian for 2D SSH model in $\bs{k}$-space is expressed by substituting the real-space fermionic operator ${c_{n_{x},n_{y}}}$ into the above Hamiltonian expression $[H_{\rm SSH}]$. For implementing periodic boundary conditions in the SSH Hamiltonian, given in Eq. \eqref{SSH_Ham}, the fermionic operator $c_{n_x,n_y}$ can be expressed in momentum space as:
\begin{equation}
c_{\bs{n}} = \frac{1}{\sqrt{N_x N_y}} \sum_{\bs{k}} e^{i \bs{n} \cdot \bs{k}} c_{\bs{k}},
\end{equation}
where \( \bs{n} = (n_x, n_y) \) and \( \bs{k} = (k_x, k_y) \), with \(N_x\) and \(N_y\) being the number of unit cells along the \(x\)- and \(y\)-directions.
The Hamiltonian in momentum space is:
\begin{equation}
H_{\rm SSH} = \sum_{\bs{k}} \Psi_{\bs{k}}^\dagger \mathcal{H}_{\rm SSH}(\bs{k}) \Psi_{\bs{k}},
\end{equation}
where \( \Psi_{\bs{k}} = \begin{pmatrix} {a}_{\bs{k}} \\ {b}_{\bs{k}} \end{pmatrix} \) is a spinor composed of sublattice annihilation operators \( {a}_{\bs{k}} \) and \({b}_{\bs{k}} \), $\bs{\sigma} = (\sigma_x, \sigma_y, \sigma_z)$ are the Pauli's pseudospin operators. The Bloch Hamiltonian $\mathcal{H}_{\rm SSH}(\bs{k})$ is written as:
\begin{subequations}
\begin{equation}
    \mathcal{H}_{\rm SSH}(\bs{k}) = \bs{h}_{\bs{k}} \cdot \bs{\sigma},
\end{equation}
where the components of $\bs{h}_{\bs{k}}= [h_x(\bs{k}), h_y(\bs{k}), h_z(\bs{k}]$ are given as:
\begin{equation}
\begin{split}
h_x(\bs{k}) &= v + w \cos k_xa + t_d^2 (1 - \cos k_xa) \cos k_ya,\\
h_y(\bs{k}) &= t_d \sin k_xa,\\
h_z(\bs{k}) &= 0,
\end{split}
\label{components}
\end{equation}
\end{subequations}
with lattice constant $a$. The eigenvalues of the Bloch Hamiltonian:
\begin{equation}
E_\pm(\bs{k}) = \pm \sqrt{h_x^2(\bs{k}) + h_y^2(\bs{k})} = \pm |\bs{h_k}|,
\label{E_pm}
\end{equation}
represent the energy bands. Substituting Eq. \eqref{components} in Eq. \eqref{E_pm}, we get:
\begin{widetext}
\begin{equation}
E_\pm(\bs{k}) = \pm \sqrt{\left[v + w \cos k_xa + t_d^2 (1 - \cos k_xa) \cos k_ya\right]^2 + t_d^2 \sin^2 k_xa}.
\end{equation}
\end{widetext}
For the 2D SSH model, where the hopping parameters alternate along different directions, the band structure shows distinct features, such as band gaps and edge states, depending on the topology of the model.

\subsection{Generalized Peierls substitution}
\label{MF}

The influence of a magnetic field perpendicular to the 2D plane of an SSH system is incorporated into the Hamiltonian \(H_{\rm SSH}\) by modifying the electronic coupling between nearest neighbors using the Peierls substitution \cite{PhysRevLett.55.2216, doi:10.1080/00107510903357796}. We incorporate the effect of the applied magnetic field along the $z$-direction in the Hamiltonian applying Peierls substitution, which modifies the hopping terms by introducing a phase factor. We consider Landau gauge along the $y$-direction, which is typically \(\mathbf{A} = (-B_zy, 0, 0)\) for a magnetic field \(\mathbf{B} = (0, 0, B_z)\). For this gauge, the phase factor for the hopping terms along the \(x\) direction is zero, and the phase factor for hopping along the \(y\) direction depends on the \(x\) coordinate. Thus, the Hamiltonian is modified as follows:
\begin{widetext}
\begin{equation}
H_{\rm M-SSH} = -\sum_{n_{x},n_{y}} \left( v \, c^{A^{\dagger}}_{n_{x}n_{y}} c^{B}_{n_{x}n_{y}} + w \, c^{B^{\dagger}}_{n_{x}n_{y}} c^{A}_{n_{x}n_{y}} + \text{H. C.} \right) 
+ \sum_{n_{x},n_{y}} \left( t_d \, c^{A^{\dagger}}_{n_{x}n_{y}} c^{B}_{n_{x},n_{y}\pm 1} e^{\pm i 2\pi \phi_{n_{x}}} + t_d \, c^{B^{\dagger}}_{n_{x}n_{y}} c^{A}_{n_{x},n_{y}\pm 1} e^{\mp i 2\pi \phi_{n_{x}}} \right),
\label{modified-SSH}
\end{equation}
\end{widetext}
where \(\phi_{n_{x}} = \frac{eB_z a^2}{h} n_x\), with \(a\) being the lattice constant and \(n_x\) the integer indexing the position along the \(x\)-axis. The parameter \(t_d\) denotes diagonal hopping strength between adjacent cells along the \(y\)-direction. This hopping parameter picks up a phase factor due to the magnetic field. The modified Hamiltonian \(H_{\rm M-SSH}\) now includes the effect of the magnetic field along the \(z\)-direction in the Landau gauge by modifying the hopping terms along the \(y\)-direction with appropriate phase factors.

\subsection{Transport properties calculation}
\label{Tp}

We calculate the transport properties of the 2D SSH by applying the non-equilibrium Green's function (NEGF) formalism, a powerful method to study quantum transport in nanostructures and open quantum systems \cite{datta2018lessons, datta2005}. It is beneficial for calculating the transmission function, which determines the probability of electron transmission through a device. The retarded Green's function \(G^{r}\) and the advanced Green's function \(G^{a}\) are defined as \cite{Datta_2005, Golizadeh_Mojarad_2007}:
\begin{equation}
G^{r}(E) = \left[EI - H_{\rm M-SSH}  - \Sigma_1(E) - \Sigma_2(E)\right]^{-1}
\label{r-g}
\end{equation}
\begin{equation}
G^{a}(E) = \left[G^{r}(E)\right]^{\dagger},
\label{a-g}
\end{equation}
where \(\Sigma_1(E)\) and \(\Sigma_2(E)\) are the self-energy matrices representing the interaction of the system with the left and right leads, respectively.

The broadening matrices \(\Gamma_1\) and \(\Gamma_2\) describe the coupling of the system to the leads and are defined as:
\begin{equation}
\Gamma_{1,2}(E) = i\left[\Sigma_{1,2}(E) - \Sigma_{1,2}^{\dagger}(E)\right].
\label{broadenning}
\end{equation}
These matrices account for the loss of electrons from the system to the leads and are crucial for calculating the transmission function. The transmission function \(T(E)\) gives the probability that an electron with energy \(E\) can pass from one lead to the other through the system. It is calculated as:
\begin{equation}
T(E) = \text{Tr}\left[\Gamma_1(E) G^{r}(E) \Gamma_2(E) G^{a}(E)\right].
\label{transmission}
\end{equation}
Here, \(\text{Tr}\) denotes the trace operation. The transmission function is a crucial quantity to determine the conductance of the system. The conductance of a quantum system can be calculated using the Landauer-B\"uttiker formula, which relates the conductance to the transmission function as
\begin{equation}
G(E) = \frac{e^2}{h} T(E),
\label{conductance}
\end{equation}
where \(e\) is the electron charge, \(h\) is Planck's constant, and \(T(E)\) is the transmission function. The local density of states (LDOS) provides information about the density of electronic states at a specific energy and position within the system, and it is given as:
\begin{equation}
\rho(\mathbf{r}, E) = -\frac{1}{\pi} \text{Im} \left[G(\mathbf{r}, \mathbf{r}, E)\right].
\label{ldos1}
\end{equation}
In the context of the NEGF formalism, the LDOS can also be expressed as:
\begin{equation}
\rho_{1,2}(E) = G^{r}(E) \Gamma_{1,2}(E) G^{a}(E).
\label{ldos2}
\end{equation}
Here, \(G^{r}(E)\) and \(G^{a}(E)\) are Green's functions as defined earlier and \(\Gamma_{1,2}(E)\) are the broadening matrices.

\section{Results}
\label{results}

\begin{figure}[t]  
\begin{tabular}{cc}	
\includegraphics[width=4.5cm,height=4.5cm]{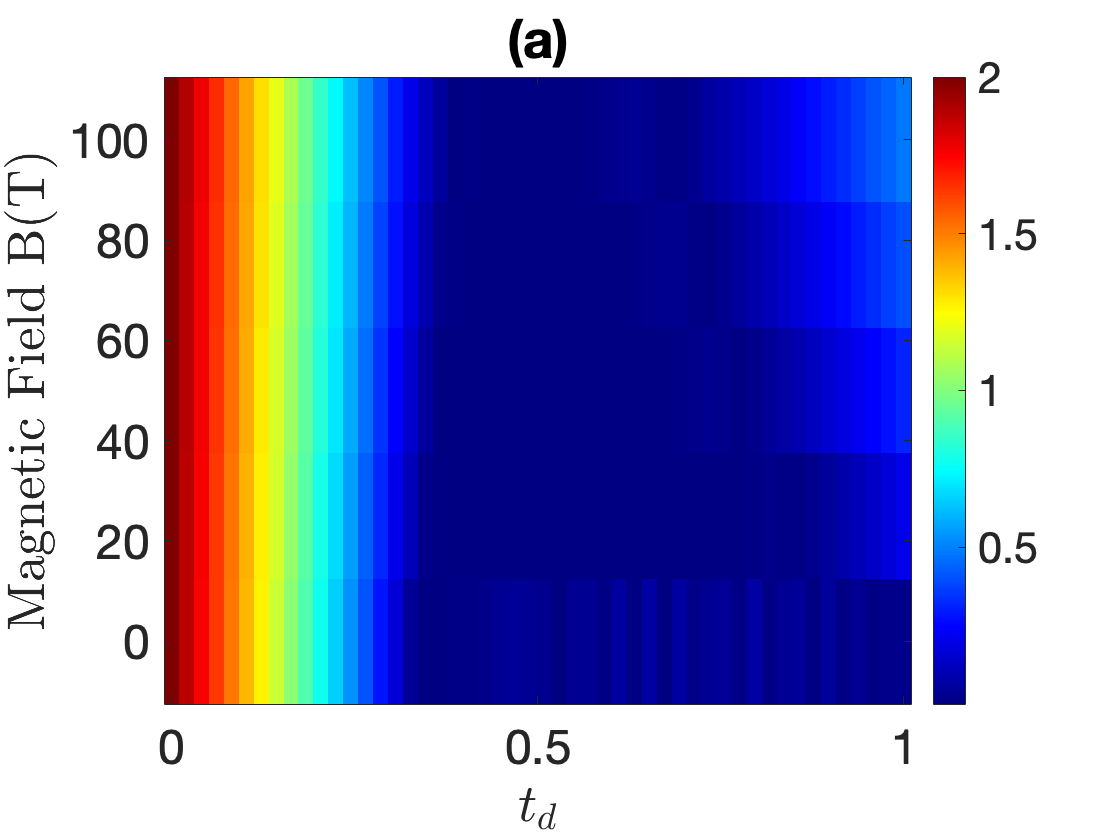}
\includegraphics[width=4.5cm,height=4.5cm]{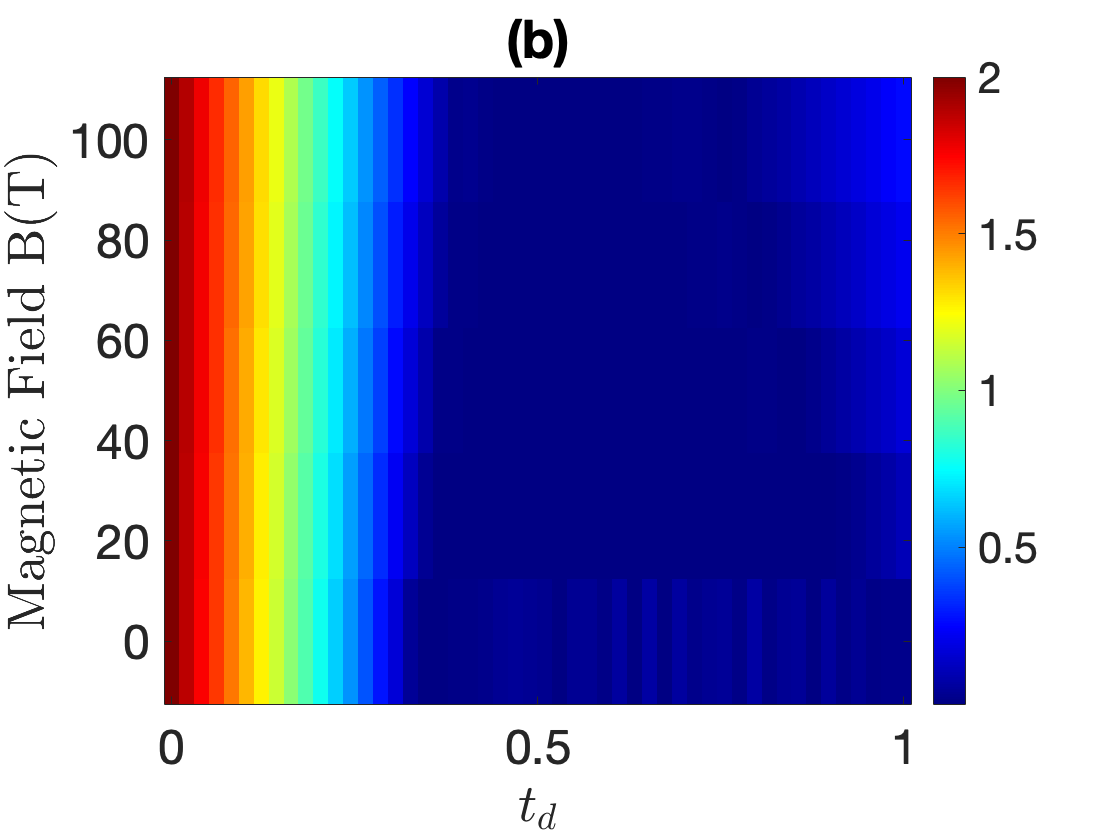}
\end{tabular}
\caption{(a) \(|v| > |w|\) and (b) \(|v| < |w|\): Color-coded energy band gap for the 2D SSH model with $N_y = 60$ dimers, illustrating the impact of varying magnetic fields from $0$ T to $100$ T and diagonal hopping parameters(\(t_d\)) ranges from $0$ to $1$, on the energy band gap for transport along the \(x\)-direction. The color scale with darker shades indicates a closed band gap, and the lighter shades represent the opening of the band gap.}
\label{Eb_M_td}
\end{figure}

\begin{figure}[b]    
\begin{tabular}{cc}	
    \includegraphics[width=4.5cm,height=4.5cm]{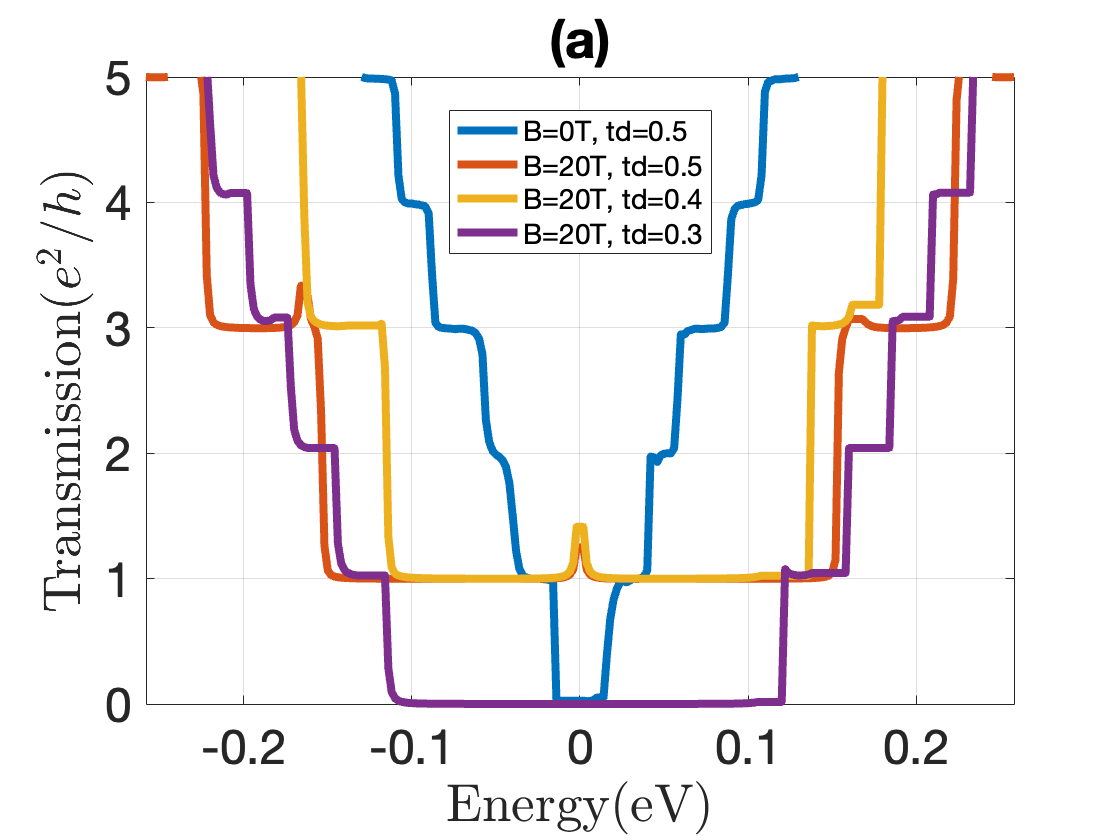}
    \includegraphics[width=4.5cm,height=4.5cm]{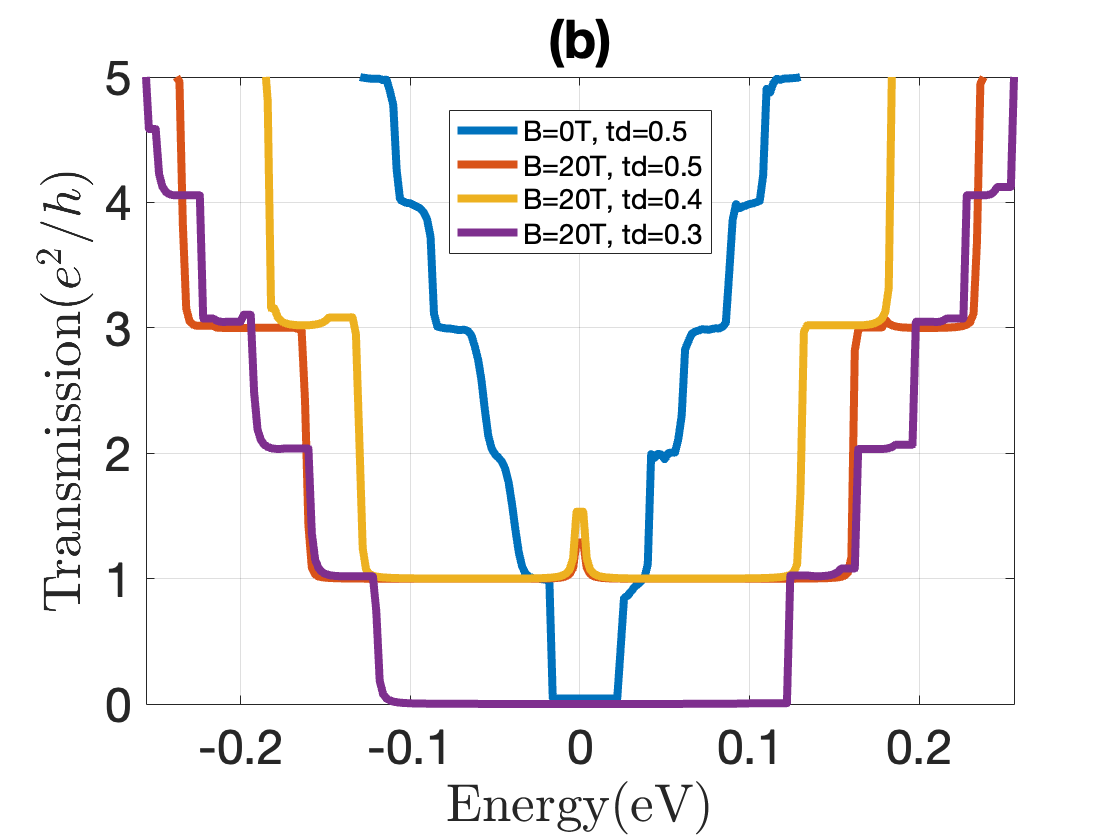}
\end{tabular}
 \caption{(a) \(|v| > |w|\) and (b) \(|v| < |w|\): Transmission(conductance: \(e^{2}/h\)) in $y$-axis as a function of energy(eV) in $x$-axis under different combination of magnetic fields strength and diagonal hopping parameter for the 2D SSH model, highlighting transitions between integer and odd integer conductance steps depending on the strength of diagonal hopping parameter(\(t_d\)).}
\label{x_transmission}
\end{figure}

\begin{figure}[t]	
\includegraphics[width=8.8cm,height=4.5cm]{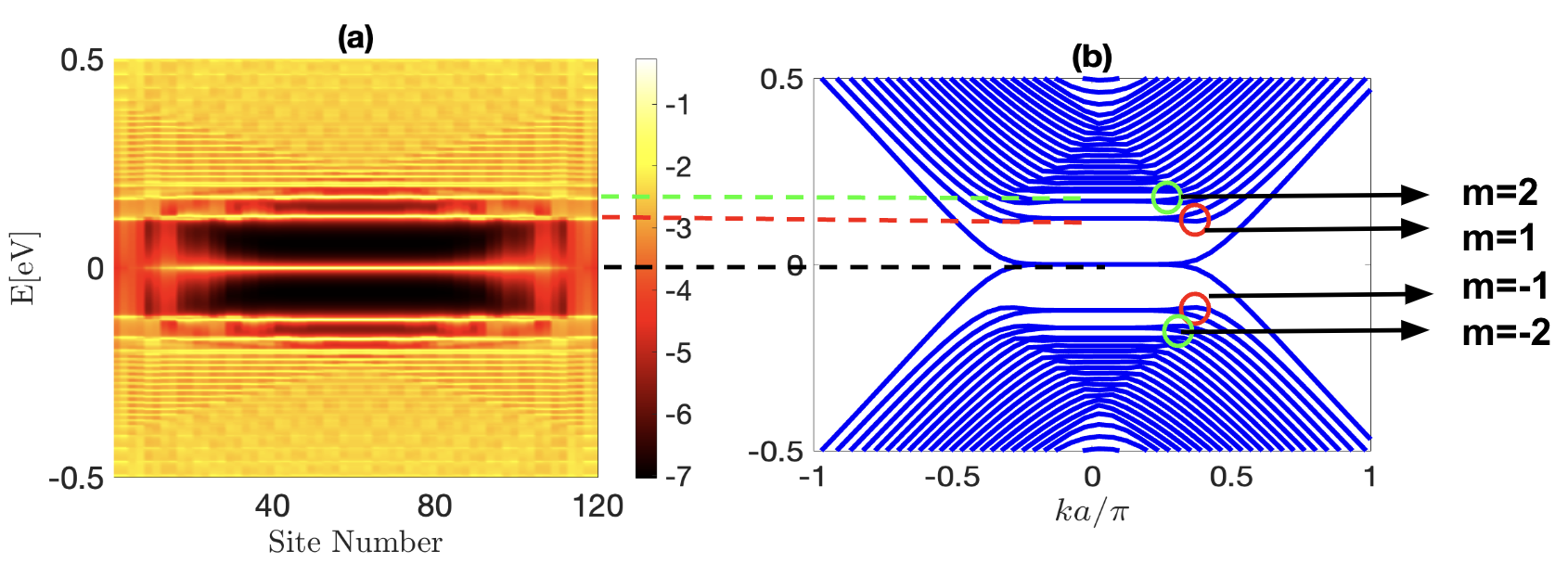}
  \caption{(a) Local density of states (LDOS) at \(B_z = 20\) T and \(t_d = 0.4\) eV, where the density of states is color-coded from lowest (black) to highest (yellow). The \(x\)-axis represents the site number, ranging from $0$ to $120$, corresponding to a 2D SSH channel of $30$ unit cells in width, with each unit cell containing four sites. The LDOS peaks at \(E = 0\) eV in the middle, indicating a high local density in the bulk. (b) Energy band diagram for the 2D SSH model in the \(|v| > |w|\) case at \(B_z = 20\) T and \(t_d = 0.4\) eV, illustrating the formation of Landau levels. The zeroth Landau level is located at \(E = 0\) eV, with the first and second Landau levels shown in green and red, respectively. The first and second levels and the higher-order Landau levels are doubly degenerate, consistent with the conductance plot for the corresponding magnetic field and \(t_d\) values.}
\label{ldos}
\end{figure}

\begin{figure}[b]
\begin{tabular}{cc}
\includegraphics[width=4.5cm,height=4.5cm]{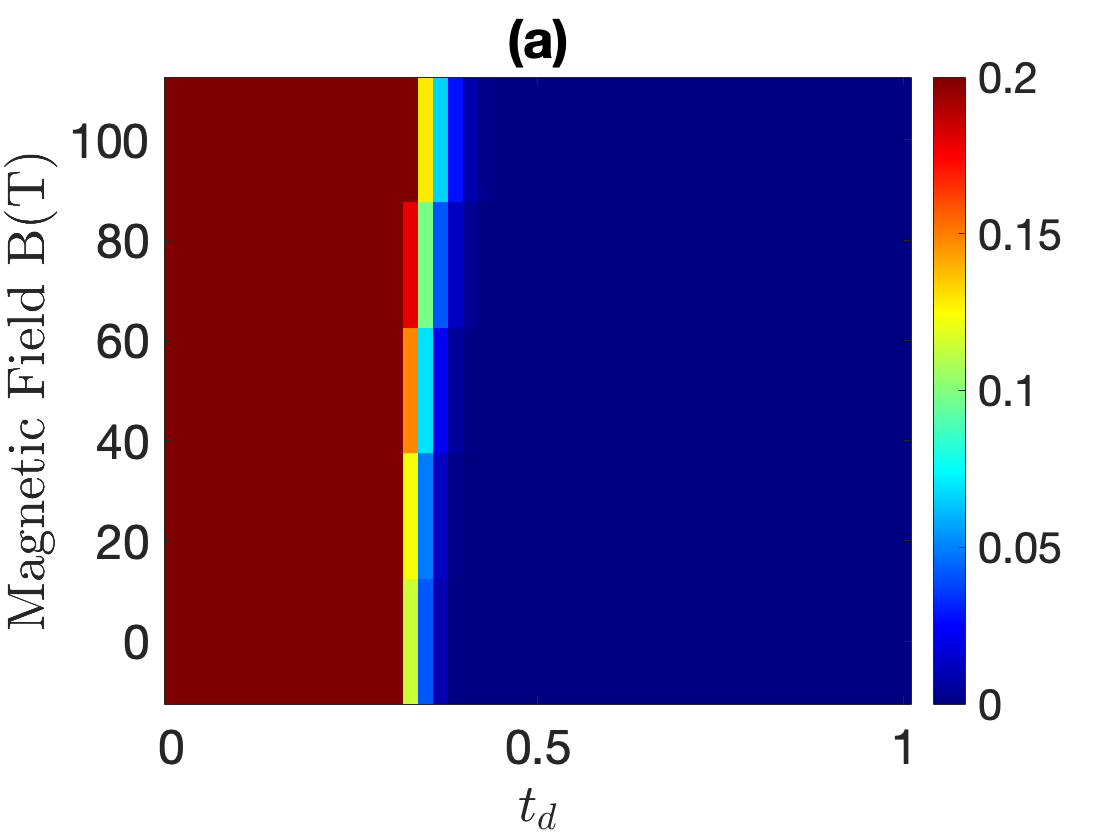}
\includegraphics[width=4.5cm,height=4.5cm]{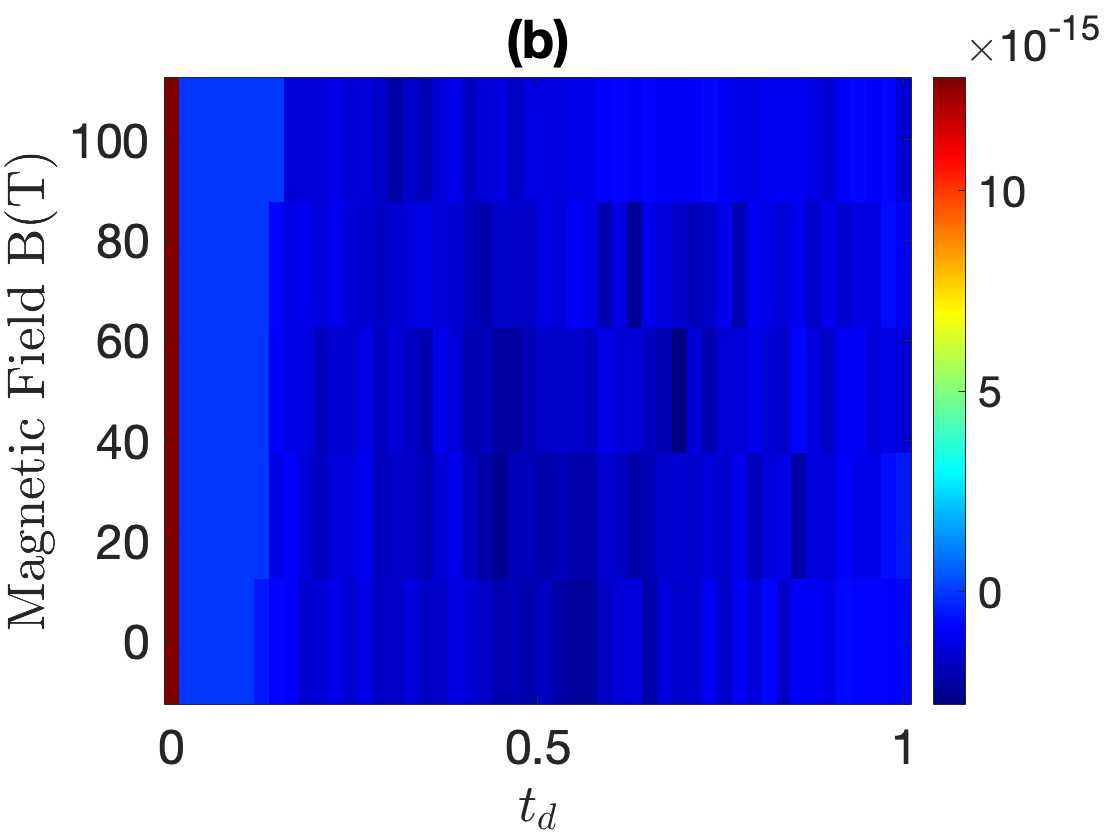}
\end{tabular}
 \caption{(a) $|v| > |w|$ and (b) $|v| < |w|$,  Color-coded energy band gap, for the 2D SSH model with $60$ dimers in width, illustrating the effect of varying magnetic fields ranging from $0$ T to $100$ T and diagonal hopping parameter (\(t_d\)) is ranging from $0$ eV to $1.0$ eV for the transport along $y$-direction.}
 \label{E_B_td_y}
\end{figure}

\begin{figure*}
\begin{tabular}{lcr}	
    \includegraphics[width=5cm,height=4.5cm]{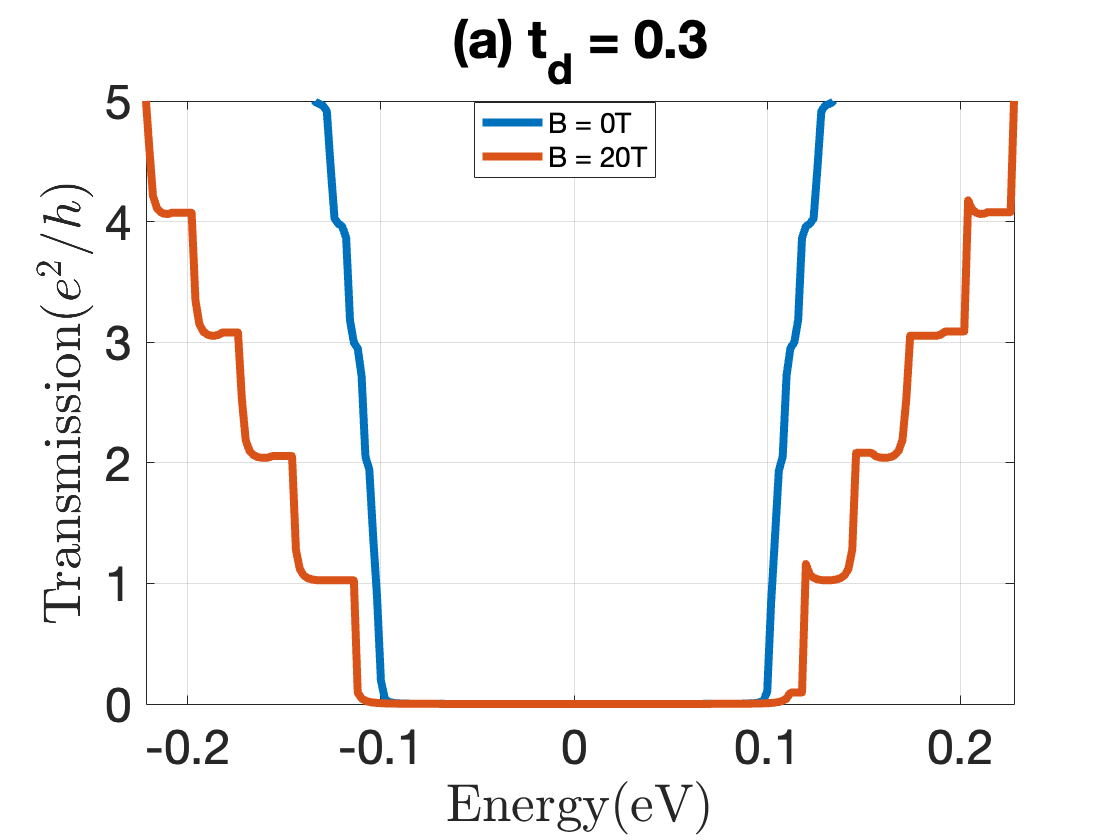}
    \includegraphics[width=5cm,height=4.5cm]{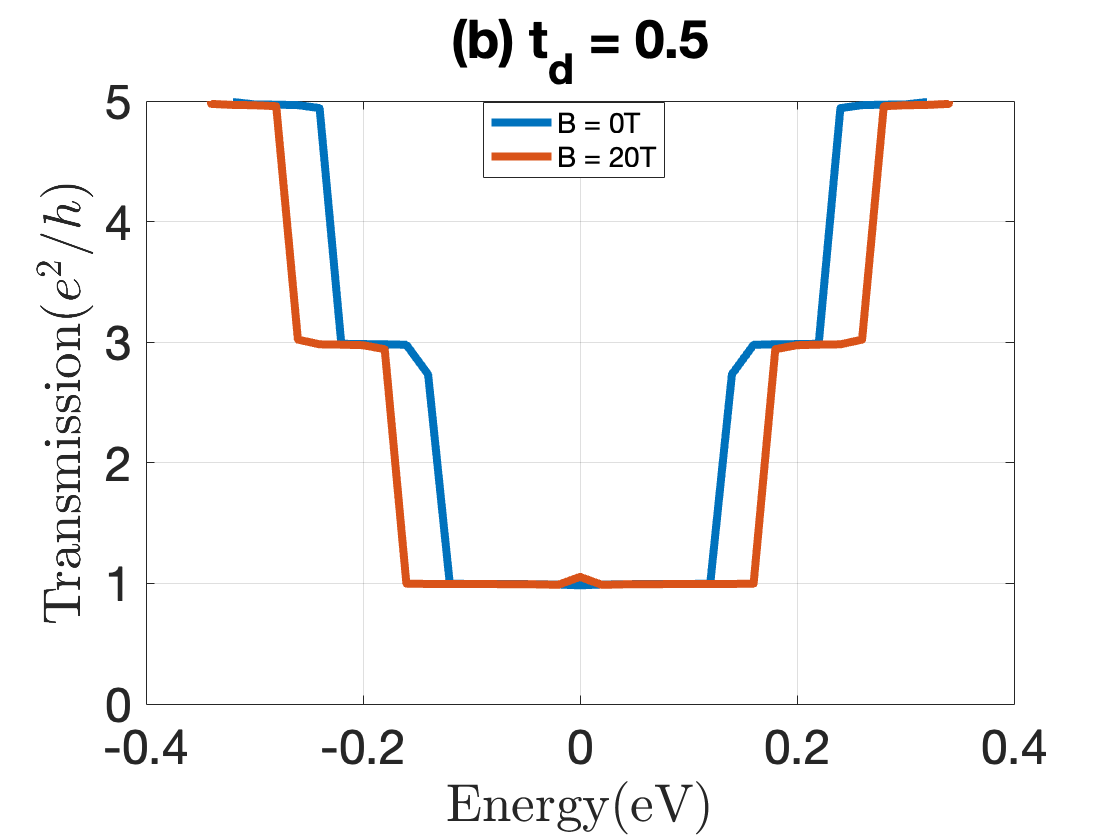}
    \includegraphics[width=5cm,height=4.5cm]{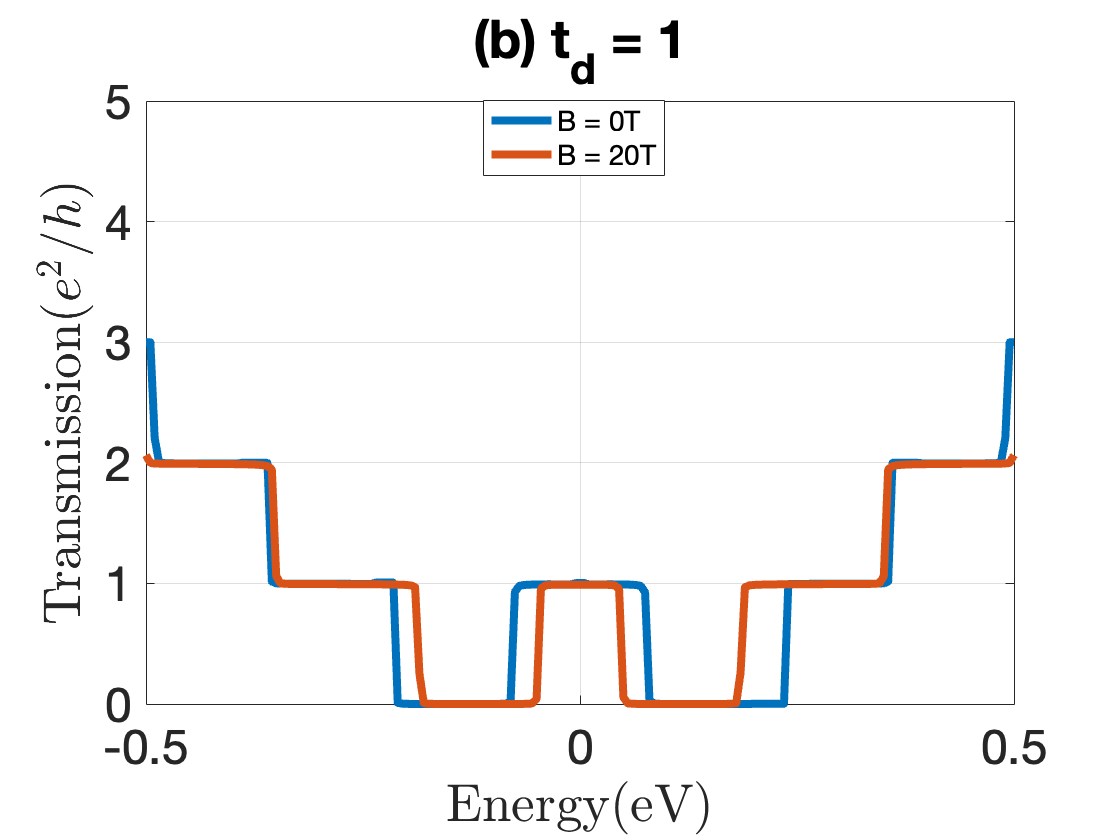} 
\end{tabular}
 \caption{Conductance(\(e^2/h\)) is plotted against energy (eV) for the 2D SSH model with (a) \(t_d = 0.3\) eV, (b) \(t_d = 0.4\) eV, and (c) \(t_d = 1.0\) eV at magnetic fields of $20$ T. Conductance steps vary between integer and odd-integer sequences based on \(t_d\) and magnetic field strength.}
 \label{tr_y}
\end{figure*}

\begin{figure}[b]
    \includegraphics[width=8.8cm,height=4.5cm]{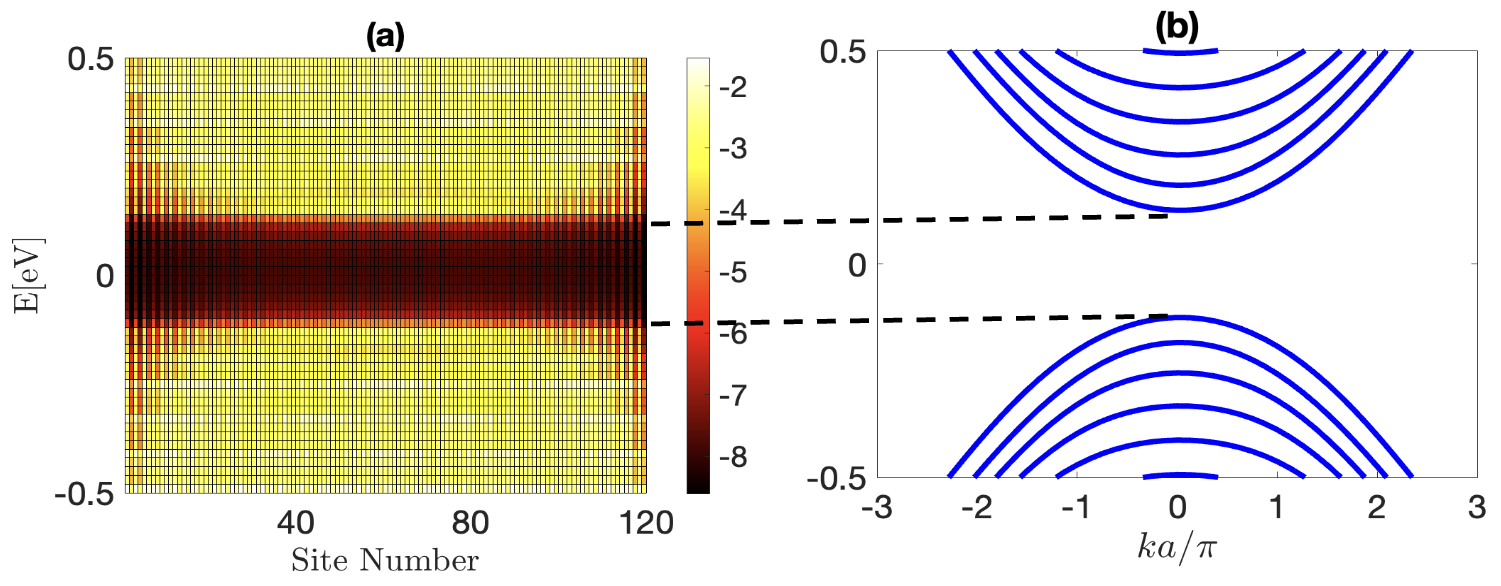} 
 \caption{(a) Local density of states (LDOS) plot at \(t_d = 0.3\) eV and \(B_z = 20\) T, with a color scale ranging from white (highest LDOS) to black (lowest LDOS). The \(x\)-axis represents the site number, spanning $0$ to $120$ across the width. (b) Band structure plot showing the energy bands as a function of the wave vector \(ka/\pi\) at \(t_d = 0.3\) eV and \(B_z = 20\) T, showing a significant band gap due to the absence of states at \(E = 0\) eV, thus indicating insulating behavior.}
\label{ldos_t=0.3}
\end{figure}

\begin{figure}[b] 
    \includegraphics[width=8.8cm,height=4.5cm]{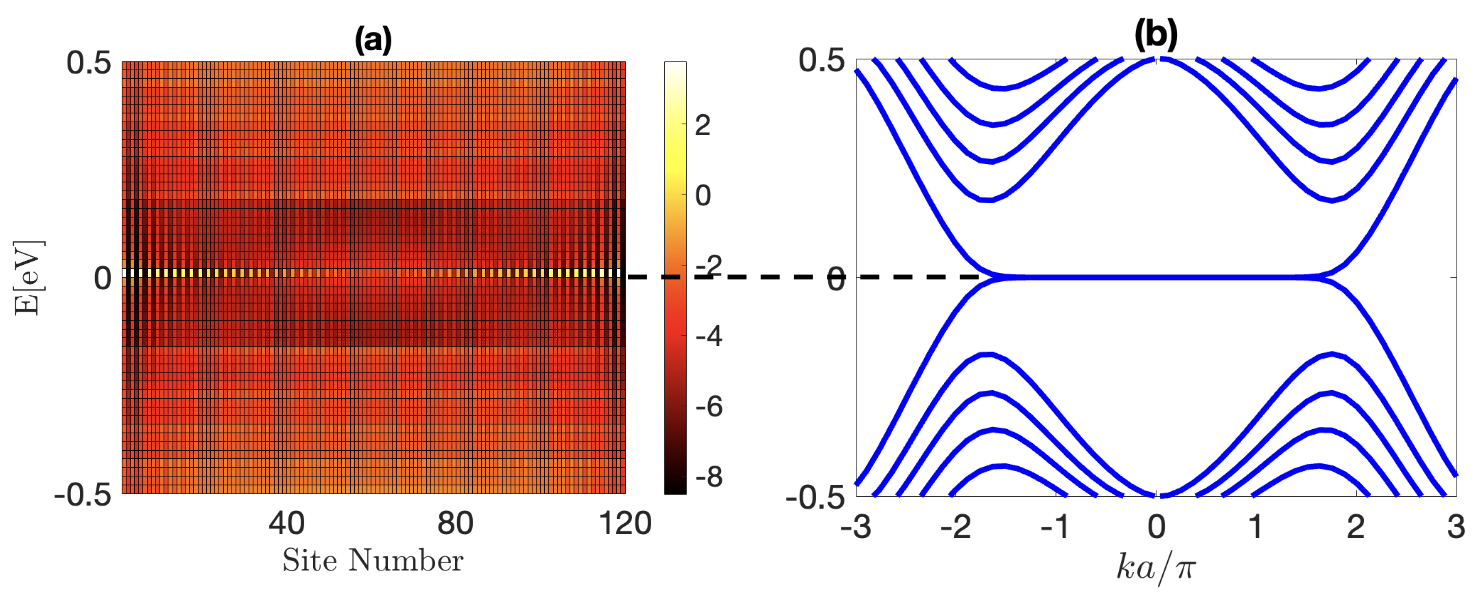}
 \caption{(a) Local density of states (LDOS) plot at \(t_d = 0.5\) eV and \(B_z = 20\) T, with a color scale ranging from white (highest LDOS) to black (lowest LDOS). The \(x\)-axis represents the site number, spanning $0$ to $120$ across the width. The high density of states can be seen at the edges of the width. (b) Band structure plot showing the energy bands as a function of the wave vector \(ka/\pi\) at \(t_d = 0.5\) eV and \(B_z = 20\) T, indicating edge conduction due to the presence of states at \(E = 0\) eV. The black dotted line indicates the states that zero energy states are responsible for the edge conduction.}
\label{ldos_t=0.5}
\end{figure}

\begin{figure}[t]
    \includegraphics[width=8.8cm,height=4.5cm]{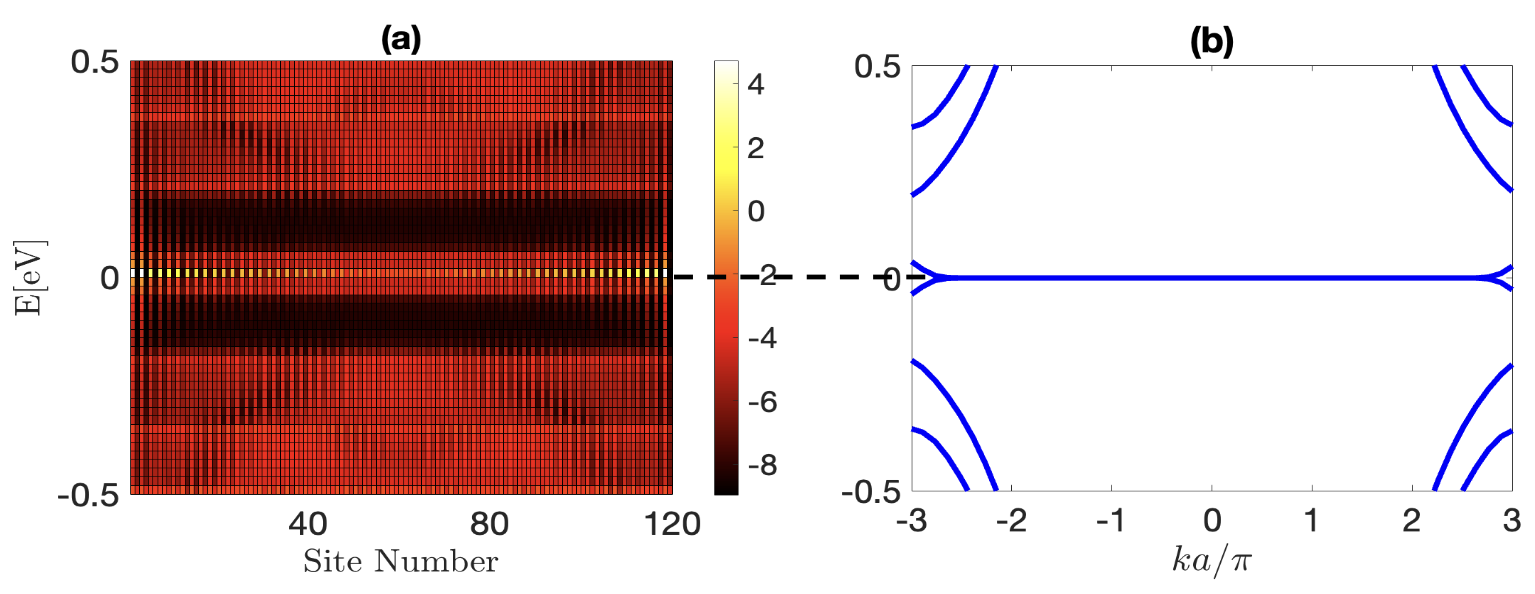}
 \caption{(a) LDOS and (b) Band-structure for \(t_d = 1.0\) eV at \(B_z = 20\) T, edge conduction persists with high state density at the system's edges.}
\label{ldos_t=1}
\end{figure}

First, this section studies the transport properties along the \(x\)-direction in the 2D SSH model. The contacts (left and right leads) are attached to the 2D SSH channel as given in Fig. \ref{Schematic}, and the electrons flow in the \(x\)-direction. The system has a width of $60$ dimers, corresponding to $120$ lattice points, and is analyzed. Subsequently, we analyze the transport properties along the \(y\)-direction in the 2D SSH system, maintaining the same width of $60$ dimers.

\subsection{Transport along $x$-direction}

 Fig. \ref{Eb_M_td}(a) presents the case where the intracell hopping (\(v\)) is greater than the intercell hopping (\(w\)), while Fig. \ref{Eb_M_td}(b) shows the scenario where \(|v| < |w|\), with a magnetic field is applied perpendicular to the system. The energy band gap, defined as the gap between the conduction and valence bands, is calculated for varying magnetic fields ranging from $0$ to $100$ T. At the same time, the diagonal hopping parameter (\(t_d\)) is varied from $0$ eV to $1.0$ eV. It is observed that the energy band gap remains zero when the magnetic field reaches $20$ T or higher, provided that the diagonal hopping parameter (\(t_d\)) exceeds $0.4$. As the magnetic field and diagonal hopping parameter increase, the band gap opens again. However, the band gap remains close when the diagonal hopping parameter strength is $t_d \simeq 0.5$ eV, and the magnetic field is above or equal to $20$ T.

We then examine the transport properties along the \(x\)-direction, specifically focusing on transmission or conductance (\(e^2/h\)) under various combinations of magnetic field strength and diagonal hopping parameters (\(t_d\)). In Fig. \ref{x_transmission}(a), the blue curve represents the transmission for \(B_z = 0\) T and \(t_d = 0.5\) eV. Here, the conductance is zero at \(E = 0\) eV, indicating a gap between the conduction and valence bands at \(E = 0\) eV, and the number of modes, or conductance steps, increases in a sequence of integer values \(n\), where \(n\) is an integer. However, for \(B_z = 20\) T and \(t_d = 0.3\) eV, depicted in magenta color, there is also no conductance at \(E = 0\) eV, and the conductance steps are increasing in the same sequence of integer values (\(n\)). For \(B_z = 20\) T and \(t_d = 0.4\) eV, colored yellow, a finite value of conduction is observed at \(E = 0\) eV, with a quantum Hall plateau from \(-0.1\) eV to \(+0.1\) eV. The number of modes or conductance steps now increases in a sequence of odd integer values, i.e., the conductance steps increase as (\(2n+1\)), where \(n\) is an integer. In the case of \(B_z = 20\) T and \(t_d = 0.5\) eV, shown in red, conduction persists at \(E = 0\) eV, with conductance steps following the same sequence of odd integers (\(2n+1\)), though the conduction window widens. Similar behavior is observed in the transmission plot for the \(|v| < |w|\) case, as shown in Fig. \ref{x_transmission}(b).
 
The quantum Hall effect is observed in the same system at a magnetic field of $20$ T, with the diagonal hopping parameter \(t_d = 0.4\) eV for both \(|v| < |w|\) and \(|v| > |w|\). At this magnetic field, Landau levels are formed. These Landau levels arise when the magnetic field is sufficiently strong to create cyclotron orbits within the system's dimensions. A wider system requires a smaller magnetic field to form these Landau levels. In Fig. \ref{ldos}(a), a bright yellow color shows the high density of states, and also the energy band diagram Fig. \ref{ldos}(b) is plotted, the zeroth Landau level is located at \(E = 0\) eV depicted by black \cite{Brey_2006, Abanin_2006}. As we move to higher energy levels, the first Landau level is depicted in green $(m=1)$, followed by the second Landau level in red. The first and second Landau levels $(m=2)$ and the higher-order Landau levels are doubly degenerate.

Examining the transmission and density of states (DOS) together for the \(|v| > |w|\) case, we observe that the DOS peaks at \(E = 0\) eV, as shown in Fig. \ref{ldos}(a), indicating a high local density of states in bulk for \(B_z = 20\) T. The energy band diagram in Fig. \ref{ldos}(b) also gets support from the observation that the number of modes, or conductance steps, increases in a sequence of odd integer values (\(2n+1\)), as seen in Fig. \ref{x_transmission}(a) for \(B_z = 20\) T and at \(t_d=0.4\) eV. Thus, this unique, odd number sequence of Landau levels gives rise to the unconventional quantum Hall effect. This result is significant because the current in the channel depends on the number of current-carrying modes when the Fermi level is between two Landau levels. At these energies, current flows through the bulk of the channel. whereas at other values of \(t_d\) we have a conventional integer quantum Hall effect. Similar conclusions can be drawn for the \(|v| < |w|\) case.

\subsection{Transport along $y$-direction}

 We now study the transport along the $y$-direction by attaching the contacts (left and right leads) to the 2D SSH channel in the $y$-direction, which leads to the electrons flowing along the $y$-direction. In Fig. \ref{E_B_td_y}(a) for \(|v| > |w|\) and Fig. \ref{E_B_td_y}(b) for \(|v| < |w|\), the energy band gap is represented by a color scale and plotted against the magnetic field and diagonal hopping parameter(\(t_d\)). In Fig. \ref{E_B_td_y}(a), the energy band gap remains open until \(t_d = 0.4\) eV, beyond which it closes regardless of the magnetic field strength. In Fig. \ref{E_B_td_y}(b), the energy band gap is nearly zero, as indicated by the color scale showing values on the order of \(10^{-15}\) eV.

To further investigate the transport properties, we calculated the transmission or conductance(\(e^2/h\)) for different combinations of magnetic field strength and diagonal hopping parameter. For the \(|v| > |w|\) case, in Fig. \ref{tr_y}(a), transmission is plotted for \(t_d = 0.3\) eV at magnetic field strengths of $0$ T and $20$ T. In both cases, there is no conduction at \(E = 0\) eV, and the conductance steps increase in integer sequences, thus showing the conventional quantum Hall effect. In Fig. \ref{tr_y}(b) For \(t_d = 0.5\) eV, at both magnetic field strength $0$ T and $20$ T, conduction occurs at \(E = 0\) eV, with conductance steps following an odd integer sequence, thus showing the unconventional quantum Hall effect. The conduction window is wider at $20$ T. As the diagonal hopping strength increases further to \(t_d = 1.0\) eV in Fig. \ref{tr_y}(c), conduction at \(E = 0\) eV persists. However, the conductance steps revert to an integer sequence, thus reverting to the conventional quantum Hall effect. This result indicates a transition from an insulating state to a conducting state, depending on the diagonal hopping parameter (\(t_d\)). In this system with a width of $60$ dimers and a transport case in the $y$-direction, the transition point is around \(t_d = 0.4\) eV.

Continuing with the analysis, we plot the local density of states (LDOS) and the energy band structure for the cases mentioned earlier. In Fig. \ref{ldos_t=0.3}(a) and Fig. \ref{ldos_t=0.3}(b), we examine the system with \(t_d = 0.3\) eV at \(B_z = 20\) T. In Fig. \ref{ldos_t=0.3}(b), there is a significantly large gap between the conduction and valence bands, which is corroborated by Fig. \ref{ldos_t=0.3}(b), where the LDOS does not show states at \(E = 0\) eV.

Next, in Fig. \ref{ldos_t=0.5}(a) and Fig. \ref{ldos_t=0.5}(b), we consider the system with \(t_d = 0.5\) eV at \(B_z = 20\) T. The band structure in Fig. \ref{ldos_t=0.5}(b) reveals that the band gap is closed. The LDOS in Fig. \ref{ldos_t=0.5}(a) indicates a high density of states at the system's edges, shown by a bright yellow color, suggesting that current flows along these edges. These are known as edge states since they are spatially located at the edges of the system \cite{hegde2021exploringideastopologicalquantum, article3}. In this scenario, the current is carried by the \(+k\) states along one edge of the sample, with an energy corresponding to the electrochemical potential of the left contact. In contrast, the \(-k\) states propagate along the opposite edge, with energy matching the electrochemical potential of the right contact. As a result, the channel becomes ballistic, with no scattering between the counter-propagating edge states.

Similarly, in Fig. \ref{ldos_t=1}(a), we present the local density of states (LDOS), and in Fig. \ref{ldos_t=1}(b), the band structure for the system with \(t_d = 1.0\) eV and \(B_z = 20\) T. The structure of the band in Fig. \ref{ldos_t=1}(b) shows a band passing directly through the center at \(E = 0\) eV. This central band is responsible for the high density of states observed at the system's edges in the LDOS plot in Fig. \ref{ldos_t=1}(b). Consequently, under this condition, the system also remains edge-conducting.

\section{Summary}
\label{summary}

In this study, we investigated the transport properties of the 2D Su-Schrieffer-Heeger (SSH) model in both the \(x\)- and \(y\)-directions under the influence of a perpendicular magnetic field. The results reveal a complex interplay between the diagonal hopping parameter (\(t_d\)) and the magnetic field, significantly affecting the system's band structure and transport characteristics. For transport in the \(x\)-direction, the system remains insulating in both cases (\(|v| > |w|\) and \(|v| < |w|\)) when no magnetic field is applied. However, under a magnetic field of $20$ T, the energy band gap remains open until \(t_d\) reaches the value $0.4$ eV. Beyond this value, it closes and leads to a transition from an insulating to a conducting state. At this point, the system also exhibits the unconventional quantum Hall effect. Despite this, the system remains bulk-conducting, as shown in the LDOS results. 

In the \(y\)-direction, the conducting or insulating nature of the system depends solely on the \(t_d\) parameter. When \(|v| > |w|\), the energy band gap remains open until \(t_d\) reaches $0.4$ eV, after which it closes, resulting in a transition to a conducting state. In contrast, the energy band gap is nearly zero when \(|v| < |w|\), indicating metallic behavior. The presence of a magnetic field, particularly at $20$ T, further alters these properties, leading to the formation of Landau levels and the quantum Hall effect.

Analysis of transport properties, including transmission and local density of states (LDOS), reveals different conduction behaviors depending on the value of \(t_d\). For \(t_d = 0.5\) eV at \(B_z = 20\) T, the system exhibits ballistic edge conduction, characterized by a high density of states at the edges and conductance steps following a sequence of odd integers. As \(t_d\) increases, the system transitions to an entirely conducting state, with conductance steps returning to an integer sequence. These findings highlight the crucial role of diagonal hopping and magnetic fields in determining the electronic properties of the 2D SSH model, providing valuable insights for the design of topological insulators and potential applications in quantum devices.

\acknowledgments
NS acknowledges financial support from DST-SERB, India, through  a Core Research   grant CRG/2023/006155. JNB acknowledges financial support from DST-SERB, India, through a MATRICS grant MTR/2022/000691.


%

\end{document}